\RequirePackage[T1]{fontenc}
\documentclass[12pt]{article}

\usepackage[height=8.85in,width=6.25in]{geometry}

\usepackage[utf8]{inputenc}
\usepackage{amsmath}
\usepackage{amssymb}
\usepackage{mathtools}
\numberwithin{equation}{section}
\usepackage{slashed}
\usepackage{braket}
\usepackage[svgnames,dvipsnames]{xcolor}
\usepackage[colorlinks,citecolor=DarkGreen,linkcolor=FireBrick,urlcolor=FireBrick,linktocpage,breaklinks=true]{hyperref}
\usepackage{cite}
\usepackage{graphicx}
\usepackage{tikz}

\usepackage{times}
\usepackage{courier}
\usepackage{bm}
\usepackage{subfig}
\usepackage{here}

\usepackage{amsthm}
\theoremstyle{plain}

\def\bZ{\mathbb{Z}}
\def\cS{\mathcal{S}}
\def\cT{\mathcal{T}}
\def\Nequals#1{$\mathcal{N}{=}#1$}
\def\vev#1{\braket{#1}}
\def\Arf{\mathop{\mathrm{Arf}}}

\begin{document}

\begin{titlepage}

\begin{flushright}

\end{flushright}

\vskip 3cm

\begin{center}

{\Large \bfseries On the trivalent junction of 
three \\[1em] non-tachyonic
heterotic string theories }

\vskip 1cm
Yuji Tachikawa
\vskip 1cm

\begin{tabular}{ll}
 & Kavli Institute for the Physics and Mathematics of the Universe (WPI), \\
& University of Tokyo,  Kashiwa, Chiba 277-8583, Japan
\end{tabular}

\vskip 2cm

\end{center}

\noindent 
Recently, Altavista, Anastasi, Angius and Uranga discussed a method to construct
junctions and bouquets of different perturbative string theories.
Following this analysis, we here argue that 
three non-tachyonic ten-dimensional heterotic string theories
can be joined together at a nine-dimensional junction.

This is done by creating a two-dimensional  non-conformal
\Nequals{(0,1)} supersymmetric quantum field theory
with three asymptotic ends, each equipped with one of the worldsheet theories of the 
supersymmetric $E_8\times E_8$ theory,
the supersymmetric $SO(32)$ theory,
and the non-supersymmetric $SO(16)\times SO(16)$ theory, respectively.
It is actually  a special case of a more general construction
involving an arbitrary $\bZ_2$-symmetric theory $T$,
its $\bZ_2$-orbifold  $T/\bZ_2$,
and the modified $\bZ_2$-orbifold $(T\times q)/\bZ_2$, 
where $q$ is a certain $\bZ_2$-symmetric spin invertible theory.

\end{titlepage}

\setcounter{tocdepth}{2}

\section{Introduction}
\label{sec:intro}

\subsection{Objective}
In addition to the two well-known supersymmetric heterotic string theories in ten dimensions, 
with gauge groups\footnote{%
We will be cavalier about the global structures of the gauge groups in this short note.
} $E_8\times E_8$ and $SO(32)$, 
there are a number of non-supersymmetric variants, one of which is non-tachyonic \cite{Alvarez-Gaume:1986ghj,Seiberg:1986by,Dixon:1986iz}.
Together, these three exhausts non-tachyonic heterotic string theories.
The non-supersymmetric non-tachyonic theory has gauge group $SO(16)\times SO(16)$,
and has a chiral spectrum which is a formal difference of the chiral spectra
of the two supersymmetric theories.
This suggests that it might be possible to connect the three non-tachyonic heterotic string theories
at a codimension-one junction.

Very recently, Altavista, Anastasi, Angius and Uranga introduced in \cite{AAAU} 
a general method to construct such junctions, or more generally bouquets, 
of different ten-dimensional perturbative string theories,
generalizing earlier works of Hellerman and Swanson, e.g.~\cite{Hellerman:2006ff,Hellerman:2007fc,Hellerman:2007zz},
where closed string tachyon condensation was used to connect different perturbative string theories.
To construct a bouquet of string theories whose internal worldsheet conformal theories are $T_1$, $T_2$, \ldots $T_n$,
the authors of \cite{AAAU} constructed a generally non-conformal theory
such that it has a scalar degree of freedom with $n$ asymptotic ends,
over which the theories $T_{1,\ldots,n}$ are fibered.
Details of the construction are given in \cite{AAAU} and therefore we will not dwell on that here.
In particular, upgrading this non-conformal theory to a conformal worldsheet theory by 
introducing dilaton gradients and dressing the interaction terms 
is much more delicate and intricate than in the case of the exact lightlike linear-dilaton solutions of 
\cite{Hellerman:2006ff,Hellerman:2007fc,Hellerman:2007zz}. 
In this short note, nothing more will be said about this important and interesting issue,
and we will simply content ourselves by constructing a non-conformal theory
connecting various conformal worldsheet theories.

Now, the authors of \cite{AAAU} discussed various examples, and left the construction of
the trivalent junction of three non-tachyonic heterotic string theories as an exercise to the reader.
The aim of this short note is to provide an answer to this exercise.

\subsection{Strategy}
It turns out that the detailed structure of the worldsheet theories of the three heterotic theories are not actually necessary. 
What is required is the following formal property, discussed e.g.~in \cite{BoyleSmith:2023xkd},
relating the three worldsheet theories using the modern understanding of $\bZ_2$ orbifolding and fermionization \cite{Karch:2019lnn,Hsieh:2020uwb}.

Recall that a ten-dimensional heterotic string theory is supersymmetric if and only if the $c=16$ internal left-moving theory is bosonic and does not depend on the worldsheet spin structure.
Indeed, the $E_8\times E_8$ current algebra theory $T$ is bosonic.
It has a non-anomalous $\bZ_2$ symmetry which preserves $SO(16)\times SO(16)\subset E_8\times E_8$,
and the $\bZ_2$ orbifold $T/\bZ_2$ is the worldsheet theory of the supersymmetric $SO(32)$ theory.
Whenever we can take the $\bZ_2$ orbifold of a bosonic two-dimensional theory,
we can use the same $\bZ_2$ symmetry to fermionize the theory,
by considering a modified $\bZ_2$ orbifold $(T\times q)/\bZ_2$,
where $q$ is a spin $\bZ_2$-symmetric invertible phase
whose partition function on a worldsheet $\Sigma$ with the spin structure $\sigma$
and a $\bZ_2$ background field $a\in H^1(\Sigma;\bZ_2)$ is given by
$(-1)^{q_{\Sigma,\sigma}(a)}$, where $q_{\Sigma,\sigma(a)}$ is the quadratic refinement
of the intersection form $\int_\Sigma ab$ for $a,b\in H^1(\Sigma;\bZ_2)$.
This fermionized theory $(T\times q)/\bZ_2$ turns out to be the worldsheet theory of 
the non-tachyonic non-supersymmetric $SO(16)\times SO(16)$ theory,
as summarized e.g.~in \cite[Figure 3]{BoyleSmith:2023xkd}.

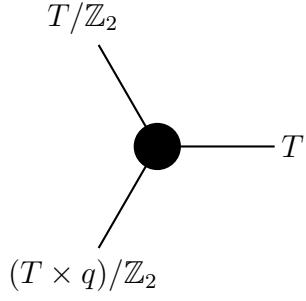
\begin{figure}
\centering
\begin{tikzpicture}[thick,scale=.6]
\def\r{2.6}
\def\rr{3}
\def\rrr{3.3}
\draw[fill=black] (0,0) circle (.5);
\draw(0,0)--(0:\r);
\node at (0:\rr) {$T$};
\draw(0,0)--(120:\r);
\node at (120:\rrr) {$T/\bZ_2$};
\draw(0,0)--(-120:\r);
\node at (-120:\rrr) {$(T\times q)/\bZ_2$};
\end{tikzpicture}
\caption{Schematic picture of the junction.
Each line represents an \Nequals{(0,1)} chiral multiplet parameterizing an asymptotic region,
together with the internal theory indicated there.
The central blob has interactions that connect  the three regions.}
\label{fig:bouquet}
\end{figure}

Therefore, our aim will be achieved if we can construct a junction of three theories,
$T$, $T/\bZ_2$ and $(T\times q)/\bZ_2$, for a general $\bZ_2$-symmetric theory $T$,
see Fig.~\ref{fig:bouquet} for a schematic picture.
As we will see shortly, we do not even have to assume that the original theory $T$ is bosonic.

\section{Construction}

Our construction is a variant of \cite{AAAU}.
In the following, we use the \Nequals{(0,1)} superspace formalism in two dimensions.

We take an arbitrary fermionic theory $T$ with non-anomalous $\bZ_2$ symmetry.
We add three chiral multiplets $X$, $\tilde X$ and $Z$
and two Fermi multiplets $\Lambda$ and $\tilde\Lambda$, 
where $X$, $Z$, $\Lambda$ are $\bZ_2$-even while $\tilde X$ and $\tilde\Lambda$ are $\bZ_2$-odd.
We gauge the total $\bZ_2$ symmetry, which is possible
since we added a $\bZ_2$-odd pair of a  left-moving fermion $\tilde\lambda$ contained in $\Lambda$ and a right-moving fermion $\psi_{\tilde X}$  contained in $\tilde X$.

We further introduce the following superpotential interaction: \begin{equation}
\int d\theta \Lambda (\tilde X^2 - X^2 - Z) +\int d\theta \tilde \Lambda X\tilde X.
\end{equation}
The classical scalar potential is \begin{equation}
V=(\tilde X^2-X^2-Z)^2 + (X\tilde X)^2
\end{equation} and the supersymmetry condition is \begin{equation}
\tilde X^2-X^2=Z, \qquad\text{and}\qquad X\tilde X=0.
\end{equation}
From this, it is easy to see that there are two classes of low-energy configurations, namely
\begin{align}
Z&\gg0, & \tilde X& =\pm \sqrt{Z}, & X=0; \label{1st}\\
Z&\ll0, & X& =\pm \sqrt{-Z}, & \tilde X=0. \label{2nd}
\end{align}

In the first case \eqref{1st}, the nonzero value of $\tilde X$ breaks the $\bZ_2$ gauge symmetry,
and the two asymptotic regions $\tilde X\to \pm\infty$ are  identified.
As the $\bZ_2$ symmetry is effectively gauge-fixed, it does not act on $T$.
There are two massive Majorana fermions of the same mass, producing no nontrivial 
invertible phases in the infrared. 
Therefore, the low-energy theory is simply a half line parameterized by $Z\gg0$,
together with its superpartner, with the original theory $T$.

In the second case \eqref{2nd}, the $\bZ_2$ gauge symmetry is unbroken,
and the two asymptotic regions $X\to \pm\infty$ are distinct.
At the same value of $Z\ll 0$, we have two points $\vev{X}=\pm \sqrt{-Z}$,
where the fermions have the interactions \begin{equation}
\lambda \psi_X \vev{X} + \tilde \lambda \psi_{\tilde X} \vev{X},
\end{equation}
where we remind the reader that the untilded fields are $\bZ_2$-even
while the tilded ones are $\bZ_2$-odd.
If we use Pauli-Villars regulator fields with large positive mass terms,
the fermions with large positive mass terms $m>0$ do not produce any phase.
In contrast, a fermion $f$ with large negative mass term $m<0$ 
produce a phase $(-1)^{\Arf(\sigma_f)}$, where $\sigma_f$ is 
the effective spin structure felt by $f$ and $\Arf$ is the Arf invariant.
In our case, we have a massive fermion $(\lambda,\psi_X)$ which \emph{does not} couple to the $\bZ_2$ gauge field $a\in H^1(\Sigma;\bZ_2)$,
and a massive fermion $(\tilde\lambda,\psi_{\tilde X})$ which \emph{does} couple to $a$.
The total invertible phase generated then has the partition function \begin{equation}
(-1)^{\Arf(\sigma+a)-\Arf(\sigma)}=(-1)^{q_{\sigma}(a)},
\end{equation} where $q_{\sigma}(a)$ is the quadratic refinement of the pairing determined
by the spin structure $\sigma$.\footnote{%
For more details of the relation between two-dimensional massive fermions, the Arf invariant $\Arf(\sigma)$ and the quadratic refinement $q_\sigma(a)$, see e.g.~\cite[Appendix A]{Karch:2019lnn} and \cite[Appendix D]{Kaidi:2019tyf}.}
Therefore, the half-line region where $Z\ll 0$ with $\vev{X}\gg0$
gives the orbifold theory $T/\bZ_2$ 
and the other region where $Z\ll0$ with $\vev{X}\ll 0$
gives the  theory $(T\times q)/\bZ_2$ obtained by first adding the invertible phase $q$
and then orbifolding the combined theory.

We thus found that our combined theory has three ends,
each described by the original theory $T$, the orbifold theory $T/\bZ_2$,
and the modified orbifold theory $(T\times q)/\bZ_2$.
This is exactly the situation depicted in Fig.~\ref{fig:bouquet}.
Finally, taking $T$ to be the $E_8\times E_8$ current algebra theory,
and the $\bZ_2$ symmetry to be the one which preserves $SO(16)\times SO(16)$ subalgebra,
we have a junction of the worldsheet theories of three non-tachyonic heterotic string theories in 
ten dimensions.

\section{Discussions}

In this short note, we constructed a two-dimensional \Nequals{(0,1)} supersymmetric theory
with three asymptotic ends, producing a junction of a theory $T$ with $\bZ_2$ symmetry,
its orbifold $T/\bZ_2$, and the modified orbifold $(T\times q)/\bZ_2$.
When $T$ and its $\bZ_2$ symmetry were appropriately chosen,
this provided a junction of the worldsheet theories of the three ten-dimensional non-tachyonic heterotic string theories.

Clearly it is of utmost importance to understand how to  promote this non-conformal theory 
into a full-fledged conformal theory describing a genuine string theory background,
and to study its detailed properties.
The nature of the junction region, where three chiral theories meet, will be of great interest.

This construction is of interest also from a more formal point of view.
An equivalence relation between two SQFTs was introduced \cite{Gaiotto:2019asa}, which roughly corresponds to 
the equivalence under continuous deformations, 
including going up and down the renormalization group flow.
By regarding the field $Z$ in our construction as non-dynamical,
we see that the limit $Z\to +\infty$ reduces to the original theory $T$,
while the limit $Z\to -\infty$ results in the direct sum of $T/\bZ_2$ and $(T\times q)/\bZ_2$.
Therefore, this provides an equivalence of the form 
\begin{equation}
T \sim  (T/\bZ_2) + ((T\times q)/\bZ_2).\label{TTT}
\end{equation}
It might also be worth mentioning that  a projective action of $SL(2,\bZ_2)$ 
on the space of $\bZ_2$-symmetric two-dimensional fermionic theories can be defined
by defining $\cS\,T:= T/\bZ_2$ and $\cT\,T=T\times q$, see e.g.~\cite{Bhardwaj:2020ymp}.
With this notation, the equivalence relation above can be more suggestively written as\footnote{%
To regard this equivalence as the one  between theories with $\bZ_2$ symmetry,
one needs to regard the left hand side as first forgetting the $\bZ_2$ symmetry of the original theory $T$, and endowing it again with a trivial $\bZ_2$ action.
This is due to the fact that the left hand side is obtained by first multiplying $T$ with a theory
with spontaneously broken $\bZ_2$ symmetry and then gauging the diagonal $\bZ_2$ symmetry.
} \begin{equation}
T \sim \cS\, T + \cS\cT\, T. \label{SSS}
\end{equation}

Finally, it is now considered more appropriate to define the equivalence relation
by utilizing  `slightly non-compact' SQFTs with asymptotic ends;
this was introduced in \cite{Gaiotto:2019gef,Johnson-Freyd:2020itv,Yonekura:2022reu}
and further discussed in \cite[Sec.~3]{Tachikawa:2025flw} and \cite[Sec.~2.2, 2.3]{Tachikawa:2025awi}.
From this perspective, the total combined theory discussed in the previous section of this note
itself defines the equivalence relation \eqref{TTT} or \eqref{SSS}.
It is widely considered quite likely that the equivalence classes of SQFTs under this relation
agree with what mathematicians call topological modular forms (TMFs).
Then the equation \eqref{TTT} or \eqref{SSS} above provides a general equality concerning $\bZ_2$-equivariant TMFs.
It would be worthwhile to investigate the relation of the equality \eqref{TTT}  or \eqref{SSS}
and the structure of $\bZ_2$-equivariant TMFs recently determined in \cite{LTY}.

\section*{Acknowledgments}
The author thanks enlightening discussions with Vittorio Larotonda.
The author is supported in part by WPI Initiative, MEXT, Japan at Kavli IPMU, the University of Tokyo
and by JSPS KAKENHI Grant-in-Aid (Kiban-C), No.24K06883.

 \bibliographystyle{ytamsalpha}
 \def\arxivfont{\rm}
 \baselineskip=.95\baselineskip
\bibliography{ref}

\end{document}